\documentclass[times,twocolumn,review]{elsarticle}

\usepackage{medima}
\usepackage{framed,multirow}

\usepackage{amssymb}
\usepackage{latexsym}
\usepackage{subfig}
\usepackage[final]{changes}

\usepackage{url}
\usepackage{xcolor}

\usepackage{array}
\usepackage{hyperref}
\usepackage{booktabs,xltabular}
\usepackage[switch,modulo]{lineno}

\definecolor{newcolor}{rgb}{.8,.349,.1}

\journal{Medical Image Analysis}

\begin{document}

\verso{Kendall Schmidt \textit{et~al.}}

\begin{frontmatter}

\title{Fair Evaluation of Federated Learning Algorithms for Automated Breast Density Classification: The Results of the 2022 ACR-NCI-NVIDIA Federated Learning Challenge}%


\author[1]{Kendall Schmidt}
\author[2,3]{Benjamin Bearce}
\author[2]{Ken Chang} 
\author[1]{Laura Coombs}
\author[4]{Keyvan Farahani}
\author[5]{Marawan Elbatel}
\author[5]{Kaouther Mouheb}
\author[5]{Robert Marti}
\author[6,7]{Ruipeng Zhang}
\author[7]{Yao Zhang}
\author[6,7]{Yanfeng Wang}
\author[8]{Yaojun Hu}
\author[8,9]{Haochao Ying}
\author[8,10]{Yuyang Xu}
\author[11]{Conrad Testagrose}
\author[12]{Mutlu Demirer}
\author[12]{Vikash Gupta}
\author[13]{Ünal Akünal}
\author[13]{Markus Bujotzek}
\author[13]{Klaus H. Maier-Hein}
\author[14]{Yi Qin}
\author[14]{Xiaomeng Li}
\author[2,3]{Jayashree Kalpathy-Cramer}
\author[15]{Holger R. Roth}


\address[1]{American College of Radiology, USA}
\address[2]{The Massachusetts General Hospital, USA}
\address[3]{University of Colorado, USA}
\address[4]{National Institutes of Health National Cancer Institute, USA}
\address[5]{Computer Vision and Robotics Institute, University of Girona, Spain}
\address[6]{Cooperative Medianet Innovation Center, Shanghai Jiao Tong University, China}
\address[7]{Shanghai AI Laboratory, China}
\address[8]{Real Doctor AI Research Centre, Zhejiang University, China}
\address[9]{School of Public Health, Zhejiang University, China}
\address[10]{College of Computer Science and Technology, Zhejiang University, China}
\address[11]{University of North Florida College of Computing Jacksonville, USA}
\address[12]{Mayo Clinic Florida Radiology, USA}
\address[13]{Division of Medical Image Computing, German Cancer Research Center, Heidelberg, Germany}
\address[14]{Electronic and Computer Engineering, Hong Kong University of Science and Technology, China}
\address[15]{NVIDIA, USA}

\received{\today}

\begin{abstract}
The correct interpretation of breast density is important in the assessment of breast cancer risk. AI has been shown capable of accurately predicting breast density, however, due to the differences in imaging characteristics across mammography systems, models built using data from one system do not generalize well to other systems. Though federated learning (FL) has emerged as a way to improve the generalizability of AI without the need to share data, the best way to preserve features from all training data during FL is an active area of research. To explore FL methodology, the breast density classification FL challenge was hosted in partnership with the American College of Radiology, Harvard Medical Schools’ Mass General Brigham, University of Colorado, NVIDIA, and the National Institutes of Health National Cancer Institute. Challenge participants were able to submit docker containers capable of implementing FL on three simulated medical facilities, each containing a unique large mammography dataset. The breast density FL challenge ran from June 15 to September 5, 2022, attracting seven finalists from around the world. The winning FL submission reached a linear kappa score of 0.653 on the challenge test data and 0.413 on an external testing dataset, scoring comparably to a model trained on the same data in a central location. 
\end{abstract}


\end{frontmatter}

\linenumbers
\nolinenumbers
\section{Introduction}
\label{sec:introduction}
Accurate assignment of breast density from screening mammograms is important for assessing breast cancer risk and future screening needs of patients as dense breasts could more easily mask tumors~\citep{Freer2015-fp}. However, this task is associated with high inter-rater variability~\citep{Sprague2016-gm}. Though AI models have been shown to learn to classify breast density with accuracy levels as high as radiologists~\citep{Chang2020-hb,Lehman2019-rk}, mammograms generated from imaging systems or patient populations outside of those used for AI training can lead to poor performance. This is especially crucial as imaging technology continues to evolve and models are deployed in regions beyond the large academic centers where deep learning models are typically developed~\citep{Kaushal2020-aq}.

Due to the sensitive nature of medical data, federated learning (FL) across a distributed data network has emerged as a popular approach to expand the diversity of training data in hopes of decreasing bias and enhancing generalizability~\citep{Rieke2020-qg}. During FL, local AI training is performed at multiple facilities allowing data to stay on-site. Only information about model updates leave facilities and are combined to create a single federated model.

FL models have been shown to be more generalizable than models trained on single institution data in medical use cases such as breast density classification~\citep{Roth2020-vb}, COVID-19 treatment prediction~\citep{Flores2021-nd}, and glioblastoma segmentation~\citep{Pati2023-cl}. Despite the promise FL holds for medical AI, the best approaches for model training (e.g., the use of batch vs. group vs. instance normalization), model aggregation (e.g., FedAvg, FedProx etc) and workflow (e.g. federated averaging vs. cyclical weight transfer vs. split learning) are active areas of research~\citep{Kairouz2019-wa}.

The recent Federated Tumor Segmentation (FeTS) FL challenge focused on optimizing aggregation method by holding the model architecture and FL workflow constant across submissions~\citep{Pati2021-qk}. The ACR-NCI-NVIDIA breast density federated learning challenge was proposed to allow participants to explore all aspects of FL methodology without constraints while also allowing participants to implement FL algorithms that work in a real distributed environment. Here, we present on the final algorithms of the challenge participants and implications for the future of medical FL. Further, we compare the performance of the FL challenge generated models with a model trained on the same clinical trial data in a central location on an external dataset and analyze their performance on different ethnicities and age groups.

\begin{figure*}[htbp]
    \centering
    \subfloat[Almost entirely fat]{\includegraphics[height=0.2\textwidth]{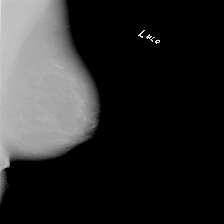}}
    \subfloat[Scattered fibroglandular densities]{\includegraphics[height=0.2\textwidth]{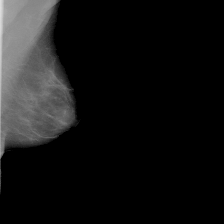}}
    \subfloat[Heterogeneously dense]{\includegraphics[height=0.2\textwidth]{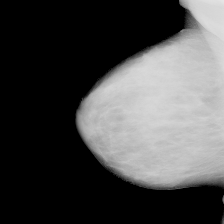}}
    \subfloat[Extremely dense]{\includegraphics[height=0.2\textwidth]{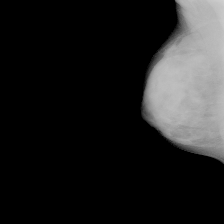}}
    \caption{Mammography examples representing each of the four BIRADS breast density categories. \label{fig:breast}}
\end{figure*}

\section{Methods}
\label{sec:methods}
\subsection{Challenge Overview}
The breast density FL challenge was hosted in a siloed environment in the American College of Radiology (ACR) Azure cloud. Interested participants could view all challenge rules and request to be registered from the publicly facing challenge website\footnote{\url{https://breastdensityfl.acr.org}}. After approval from a challenge admin, participants were given access to a private challenge URL through which challenge submissions could be made. 
Participants were tasked with developing FL algorithms which could accurately classify breast density, examples in (Fig.~\ref{fig:breast}). Participants were allowed to use pretrained starting models. Organizers did not participate in the challenge. 

Submissions for this challenge were zip archives consisting of instructions to build docker containers which could launch automated FL runs. The challenge was broken up into two phases. Phase I, which ran from June 15th 2022 to September 5th 2022, was for training and developing the FL algorithms. During Phase I, participants were limited to three submissions per day. Phase II consisted of running each participant’s last Phase I submission on a hold-out test dataset. At the end of Phase I, all active participants were given a questionnaire on their demographics, affiliations, and FL methodology. The results of the challenge were announced at MICCAI 2022, Singapore with the winning submission receiving a GPU prize from NVIDIA.

\subsection{Challenge Architecture}
\begin{figure}[htbp]
    \centering
    \includegraphics[width=0.95\columnwidth]{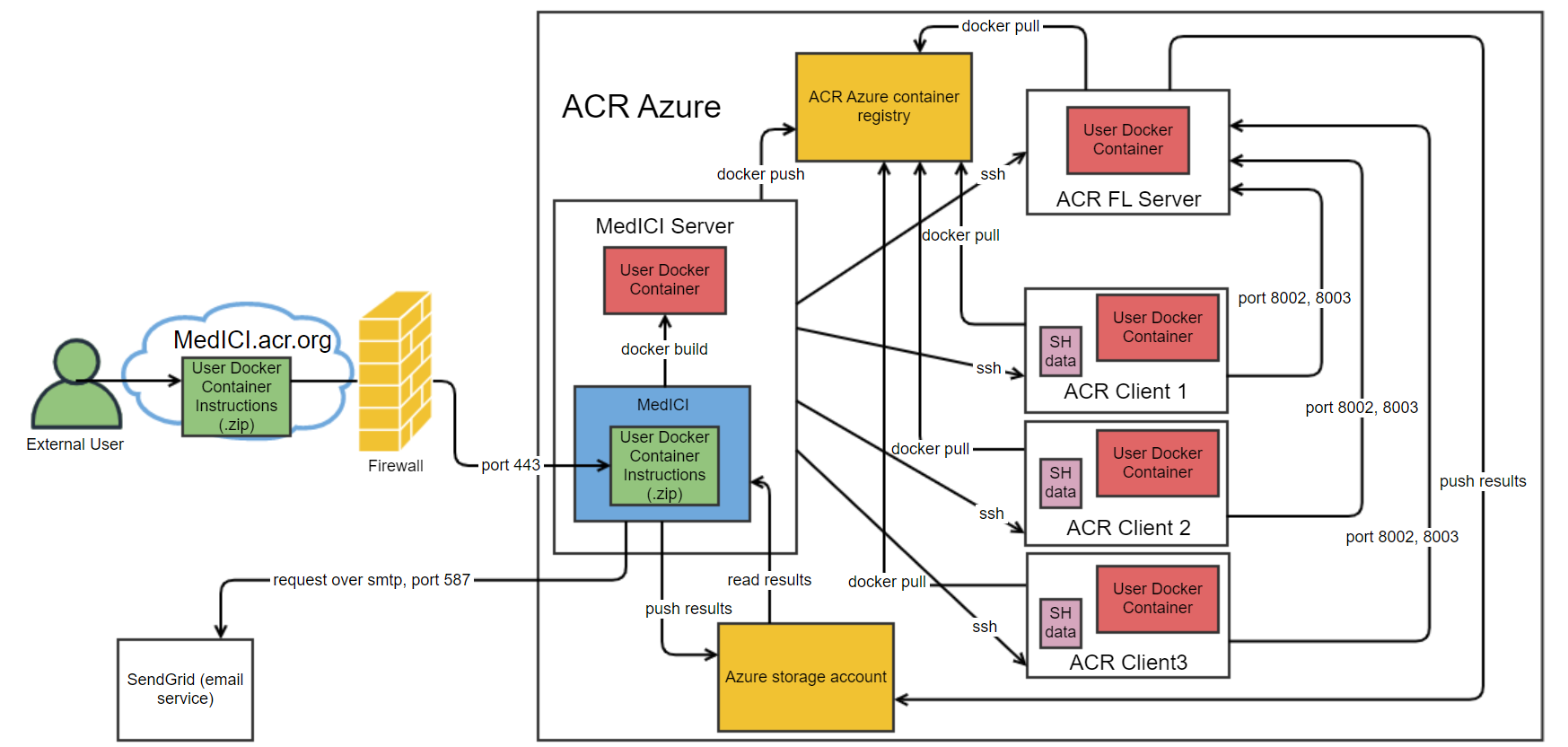}
    \caption{Breast density federated learning challenge architecture diagram: Depicted inside the white ACR Azure square are 5 smaller white VM squares. These are from left to right the MedICI Server, FL Server, and 3 FL clients. User containers are in red. Other resources include Azure storage (bottom yellow), Azure docker image container registry (top yellow), data indicated by “SH data” purple squares, uploaded algorithms as .zip in green. Finally a firewall was used to control access. After approval by challenge admin, users were allowed access to submit their FL algorithms through MedICI, which coordinated the kickoff of FL runs across the ACR FL server and the three ACR Clients.  \label{fig:architecture}}
\end{figure}

\added{The Challenge architecture consisted of five virtual machines (Fig.~\ref{fig:architecture}). One ran Medical Imaging Challenge Infrastructure (MedICI) software to receive and distribute challenge submissions. One served as an FL server to coordinate FL runs across clients. The remaining three were FL clients each with unique large breast density datasets for model training.} 
\subsubsection{\added{MedICI}}

Challenge participants interfaced with the challenge environment via \replaced{MedICI}{Medical Imaging Challenge Infrastructure (MedICI)}, a CodaLab-based~\citep{Pavao2022-vw} open source web platform designed to help data scientists and research teams to crowd-source the resolution of machine learning problems through the organization of competitions. MedICI integrates the job scheduling feature from CodaLab with support for running docker containers built from  uploaded algorithms and code. MedICI was developed at the Massachusetts General Hospital (MGH) with support from the National Cancer Institute. 

An instance of the MedICI website was hosted on the MedICI server (Fig.~\ref{fig:architecture}). Through the MedICI website, participants could submit zip files consisting of instructions to build their docker images. For security reasons, the automated docker build process was turned manual in order to review the uploaded docker build instructions to ensure no malicious content. After review, docker images were built on the MedICI server then manually pushed to a container repository. Once pushed, the docker images became available on the platform. 

Participants could select to submit FL algorithms contained within their docker images from a dropdown list at which point MedICI would add the submission to a queue. MedICI kicked off one submission at a time by initiating a docker pull and subsequent docker run of the selected container on the FL components (FL server and 3 FL clients). On completion of the submission or after 8 hours of runtime, MedICI terminated all running containers, initiated a push of all generated results from the FL server to a storage account, and kicked off the next submission in the queue. Results for each run were stored and displayed to users.

\subsubsection{\added{FL environment}}
An FL server coordinated FL submissions across three client machines. Each client machine had a \added{NVIDIA Tesla V100} GPU and a unique large dataset, described in more detail below. \added{Though some challenges have restricted computational resources in order to force computationally efficient submissions, that was not done here. The full use of each GPU was available to participants to allow a focus on FL methodology.} For a participant submission to be run successfully, it needed to be fully automated after docker run and complete within 8 hours of runtime. \added{The total run time was not a factor in submission ranking.} To demonstrate one such submission, a reference submission using  NVIDIA FLARE~\citep{Roth2022-ru} and MONAI~\citep{Cardoso2022-uy} was created and made publicly available\footnote{\url{https://github.com/Project-MONAI/tutorials/tree/main/federated_learning/breast_density_challenge}}. Though not a submission requirement, all submissions throughout the challenge duration used NVIDIA FLARE.

\subsection{Challenge Data}
\begin{table}[htbp]
    \caption{Number of images per data split across clients.  
    \label{tab:nr_images}}
    \centering\footnotesize
    \begin{tabular}{lllll}
    \toprule
        \textbf{} & \textbf{Site-1} & \textbf{Site-2} & \textbf{Site-3} & \textbf{Overall} \\ \hline
        \textbf{Training} & 22964 & 6534 & 40031 & 69529 \\ 
        \textbf{Test 1} & 3327 & 1128 & 5952 & 10407 \\ 
        \textbf{Test 2} & 6637 & 3889 & 13428 & 23954 \\ \hline
        \textbf{Total} & 32928 & 11551 & 59411 & 103890 \\ 
    \bottomrule
    \end{tabular}
\end{table}

Data used for this challenge were obtained as part of the Digital Mammographic Imaging Screening Trial (DMIST)~\citep{Pisano2005-zh}. ACR holds the IRB for the DMIST dataset and access must be explicitly approved. Thus, no DMIST data was shared with challenge participants. As such, a sample dataset was provided along with the reference example. The sample dataset was based on the Curated Breast Imaging Subset of Digital Database for Screening Mammography (CBIS-DDSM) dataset~\citep{Lee2017-wq} hosted by The Cancer Imaging Archive (TCIA)~\citep{Clark2013-zx}. Both the DMIST dataset and the sample dataset were preprocessed as described in~\citep{Roth2020-vb}.

\begin{figure}[htbp]
    \centering
    \includegraphics[width=0.85\columnwidth]{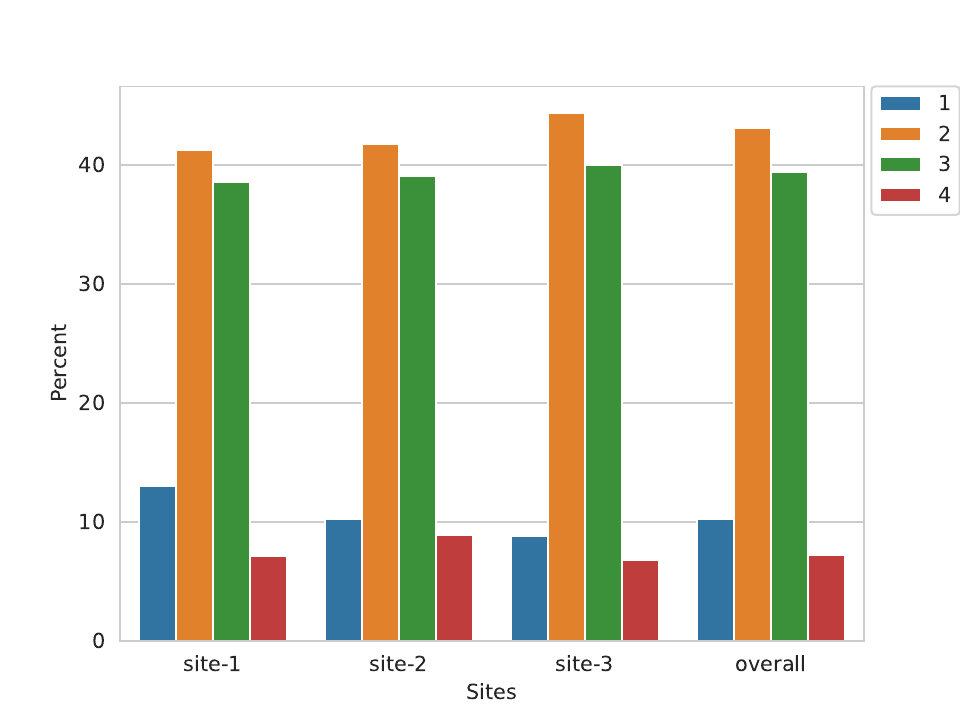}
    \caption{BI-RADS scores across sites. \added{1: Almost entirely fatty, 2: Scattered areas of fibroglandular tissue,
3: Heterogeneously dense, 4: Extremely dense.}  \label{fig:birad_scores}}
\end{figure}

DMIST contains data from 33 institutions and was first split across FL clients by digital mammography system as described here:~\citep{Chang2020-hb}. Then, all but the majority scanner type for each client split were omitted as here~\citep{Gupta2021-hs} and all patients with data in multiple splits were moved to the test 2 split. Table~\ref{tab:nr_images} summarizes the final data splits across clients for the two challenge phases. As part of DMIST, these data were labeled with one of the four ACR BI-RADS breast density categories (almost entirely fatty, scattered fibroglandular densities, heterogeneously dense, or extremely dense). Figure~\ref{fig:birad_scores} shows the distribution of BI-RADS scores across sites. Only the labels for the training data were provided to the participants. Test 1 data was used for Phase I ranking and test 2 data was used for Phase II ranking.

\subsection{Ranking}

Site-level (linear kappa, quadratic kappa, and area under the ROC curve (AUC)) and image-level (linear and quadratic distances) metrics were calculated per site and overall. These metrics are commonly used to assess breast density classification AI. While the site-level metrics provided insight on the overall accuracy of the AI, the image-level metrics provided a measure of the degree correctness as breast density is measured as an ordinal scale. To rank each submission, metrics were ranked individually and then averaged to obtain a single consensus ranking. \added{Given} 

\begin{itemize}
  \item[] \added{$M$: the number of metrics used for ranking.}
  \item[] \added{$R_{ij}$: the rank of participant $i$ for metric $j$.}
\end{itemize}

\added{The average rank $\bar{R_{i}}$ for each participant $i$ was calculated:}

\begin{equation}
    \added{      \bar{R_{i}} = \frac{1}{M} \sum_{j=1}^{M}R_{ij}    }
\end{equation}

Ranking in this way allowed overall submission accuracy and degree of correctness to be taken into account. Image-level metrics were used to perform statistical stability analysis on Phase II rankings with ChallengeR\footnote{\url{https://github.com/wiesenfa/challengeR}}~\citep{Wiesenfarth2021-bj}.

\section{Post-Challenge Analysis}
\label{sec:post-challenge}
\subsection{External Validation}
All Phase II algorithms were validated on an external test set from MGH. This evaluation was performed outside the MedICI platform on compute resources at MGH. The dataset was obtained through IRB approval with patients who had prior surgery or implants excluded. Each image was read by a breast imaging radiologist as part of routine clinical practice. All mammograms were acquired using a Lorad Selenia mammography system (Hologic). The final MGH patient cohort consisted of 8,603 digital screening images from 1,856 patients.  Since all participants used NVIDIA FLARE, a custom script was created to call the Learner class for each method and run inference on this novel dataset. Linear kappa, quadratic kappa and AUC metrics were compared between the Phase II results and the external validation.

Previously, a model had been trained on DMIST data in a non-federated manner using similar data splits~\citep{Chang2020-hb}. We refer to this model as the central DMIST model. The central DMIST model was also evaluated against the external validation dataset. Linear distance and quadratic distance were calculated per image and compared to the FL methods. Since, for the challenge, patients with data overlap across multiple splits were moved to the test 2 split, the test 2 split contained some data used to train the central DMIST model. Therefore, this model was not evaluated on the test 2 split.

\subsection{Demographic Bias Analysis}
\begin{table}[htbp]
    \caption{Patient race and age distribution across the different testing sites.  \label{tab:race_age}}
    \centering\footnotesize
    \begin{tabular}{m{7em}llll}
    \toprule
        \textbf{Race} & \textbf{Site-1} & \textbf{Site-2} & \textbf{Site-3} & \textbf{Overall} \\ \hline
        American Indian or Alaska & 0 & \added{0} & \added{2} & \added{2}\\ 
        Asian & 33 & 13 & 37 & 83 \\ 
        Black or African American & 176 & 109 & 389 & 674 \\ 
        Hispanic or Latino & 41 & 62 & 72 & 175 \\ 
        Other & 10 & 15 & 18 & 43 \\ 
        Unknown & 5 & 0 & 0 & 5 \\ 
        White & 1122 & 563 & 2237 & 3922 \\ \hline
        \textbf{Total} & 1387 & 762 & 2755 & 4904 \\ 
    \bottomrule
        \textbf{} & ~ & ~ & ~ & ~ \\ 
    \toprule
        \textbf{Age} & \textbf{Site-1} & \textbf{Site-2} & \textbf{Site-3} & \textbf{Overall} \\ \hline
        \textbf{Mean +- Std. Dev.} & 55.2+-10.7 & 53.1+-9.8 & 54.2+-10.2 & 54.3+-10.3 \\ 
        \textbf{Range} & 23-90 & 29-89 & 24-88 & 23-90 \\ 
    \bottomrule
    \end{tabular}
\end{table}
To assess how different patient demographics can influence the performance of the finalist methods on the Phase 2 test dataset, for each submission, per-image metrics were examined along race and age (rounded to the nearest 10 years) categories. Since DMIST data consists of all female subjects, gender bias was not a concern. Table~\ref{tab:race_age} summarizes the patient self-identified race and age data across the different sites.  

\section{Results}
\label{sec:results}
\subsection{\added{Phase II submissions}}
\begin{table*}[htbp]
    \caption{Phase II submission summary.  \label{tab:phase2_summary}}
    \centering\footnotesize
    \begin{tabular}{p{1.5em}p{1.5em}p{18.5em}p{4em}p{6em}p{6em}p{6em}p{6em}}
    \hline
        \textbf{Rank} & \textbf{Algo.} & \textbf{Team} & \textbf{Country} & \textbf{Base model architecture} & \textbf{General-Purpose pre-trained weights} & \textbf{Task-specific pre-trained weights} & \textbf{FL algorithm} \\ \hline
        \textbf{1} & \#1 & Elbatel, Mouheb, Marti & Spain & ResNet50  & No & Yes & FedProx \\ 
        \textbf{2} & \#2 & Zhang, Zhang, Wang & China & ResNet50  & Yes & No & FedAvg \\ 
        \textbf{3} & \#3 & Hu, Ying, Xu & China & DenseNet-121 & Yes & No & SCAFFOLD \\ 
        \textbf{4} & \#4 & Testagrose, Demirer, Gupta, Liu, Erdal, White & USA & Inception-V3  & Yes & Yes & FedAvg \\ 
        \textbf{5} & \#6 & Akünal, Bujotzek, Maier-Hein & Germany & ResNet18 & Yes & No & FedAvg \\ 
        \textbf{6} & \#5 & Bahrini & Tunesia & ResNet50 & Yes & No & FedAvg \\ 
        \textbf{7} & \#7 & Qin, Li & China & ResNet56 & No & No & FedAvg \\ \hline
    \end{tabular}
\end{table*}
\added{A summary of the Phase II submissions based on the participant questionnaire and provided abstracts is shown in Table~\ref{tab:phase2_summary}. For each submission, a brief summary is provided below, with more detailed summaries available in ~\ref{sec:abstracts}. Note these summaries were provided by the participants. Challenge organizers had limited insight into the methods beyond what was provided by participants.}

\subsubsection{\added{1st Rank (Algo. \#1)}}
\added{This method began with normalizing the imaging data histograms. A ResNet50 architecture pre-trained on task-specific data was used along with a rank-consistent layer for ordinal classification. For 100 rounds, FedProx~\citep{Li2018-xf} was used after 1 epoch. The classification layer was not globally aggregated, thus each site maintained a locally fine-tuned classification layer. For external validation, the three local models generated during FL were treated as an ensemble.}
\subsubsection{\added{2nd Rank (Algo. \#2)}}
\added{This method used a ResNet50 architecture pretrained on generic image data. Like the 1st rank algorithm, this method also maintained local classifiers at each site, however, a global classifier was also maintained. Model aggregation occurred via FedAvg~\citep{McMahan2017-tc} for 80 rounds each after two local training epochs.}
\subsubsection{\added{3rd Rank (Algo. \#3)}}
\added{A DenseNet121 architecture was used in this method with SCAFFOLD~\citep{Karimireddy2019-bi} for model aggregation. This method also employed Top-k sparsification perhaps due to communication bandwidth issues experienced with prior submissions.}
\subsubsection{\added{4th Rank (Algo. \#4)}}
\added{For this method the Inception-V3~\citep{Szegedy2016-mq} architecture pre-trained on generic image data was used. Prior to submitting to the challenge, the model was fine-tuned on task specific data from the developer’s facility. The method employed FedAvg for model aggregation.}
\subsubsection{\added{5th Rank (Algo. \#6)}}
\added{A ResNet18~\citep{he2016deep} model architecture was used with FedAvg for aggregation after 1 epoch for 100 rounds. FedProx was also tested in this method but ultimately found to be less accurate than FedAvg. Other model architectures were tried, but issues were faced in terms of finishing prior to the 8 hour time limit or with communication bandwidth.}
\subsubsection{\added{6th Rank (Algo. \#5)}}
\added{This participant did not provide a description of this method.}
\subsubsection{\added{7th Rank (Algo. \#7)}}
\added{A ResNet56 model was used for this method. Aggregation occurred via FedAvg, weighted by the proportion of total data points from each site. Between each round, each site trained for 5 epochs.}
\subsection{Challenge evaluation}
\deleted{A summary of the Phase II submissions based on the participant questionnaire and provided abstracts is shown in Table~\ref{tab:phase2_summary}.}
\begin{table*}[htbp]
    \caption{Average performance across final algorithms.  \label{tab:avg_scores}}
    \centering\footnotesize
    \begin{tabular}{p{8em}p{3em}p{3em}p{3em}p{3em}p{3em}p{3em}p{3em}p{3em}p{3em}p{3em}p{3em}p{3em}}
    \toprule
        \textbf{} & \textbf{Lin. Kappa Site-1} & \textbf{Quad. Kappa Site-1} & \textbf{AUC Site-1} & \textbf{Lin. Kappa Site-2} & \textbf{Quad. Kappa Site-2} & \textbf{AUC Site-2} & \textbf{Lin. Kappa Site-3} & \textbf{Quad. Kappa Site-3} & \textbf{AUC Site-3} & \textbf{Lin. Kappa All} & \textbf{Quad. Kappa All} & \textbf{AUC All} \\ \hline
        \textbf{Phase I (Train \& Val.)} & 0.59 & 0.69 & 0.9 & 0.49 & 0.59 & 0.92 & 0.65 & 0.75 & 0.93 & 0.62 & 0.72 & 0.91 \\ 
        \textbf{Phase II (Test)} & 0.63 & 0.73 & 0.93 & 0.47 & 0.58 & 0.90 & 0.66 & 0.75 & 0.94 & 0.62 & 0.72 & 0.93 \\ \hline
        \textbf{Relative Change} & 7.42\% & 5.79\% & 2.81\% & -4.62\% & -1.40\% & -2.24\% & 0.64\% & 0.55\% & 0.85\% & 0.93\% & 0.90\% & 1.28\% \\ 
    \bottomrule
    \end{tabular}
\end{table*}

The average performance of the final algorithms across evaluation metrics between Phase I and Phase II is shown in Table~\ref{tab:avg_scores}. On average, the final algorithms performed slightly better on the Phase II data but show a relatively stable performance across the two phases of the challenge.
\begin{figure*}[htbp]
    \centering
    \subfloat{\includegraphics[height=0.65\columnwidth]{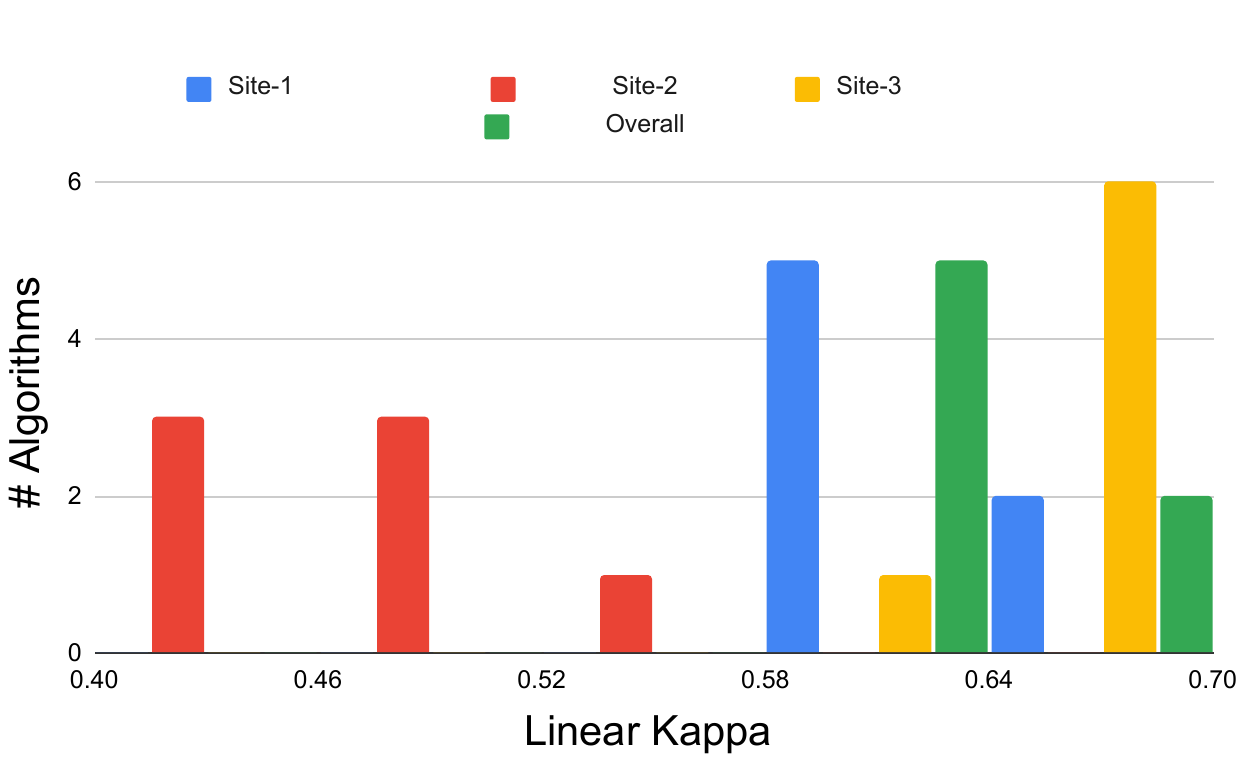}}
    \subfloat{\includegraphics[height=0.65\columnwidth]{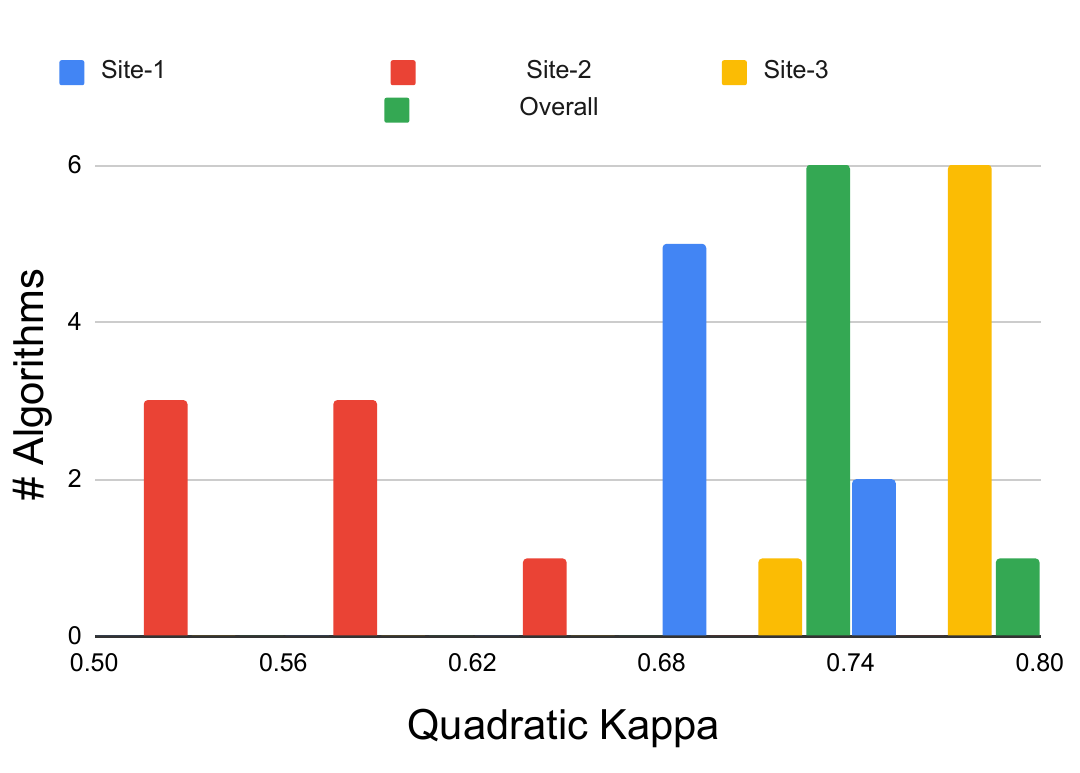}}\\
    \subfloat{\includegraphics[height=0.65\columnwidth]{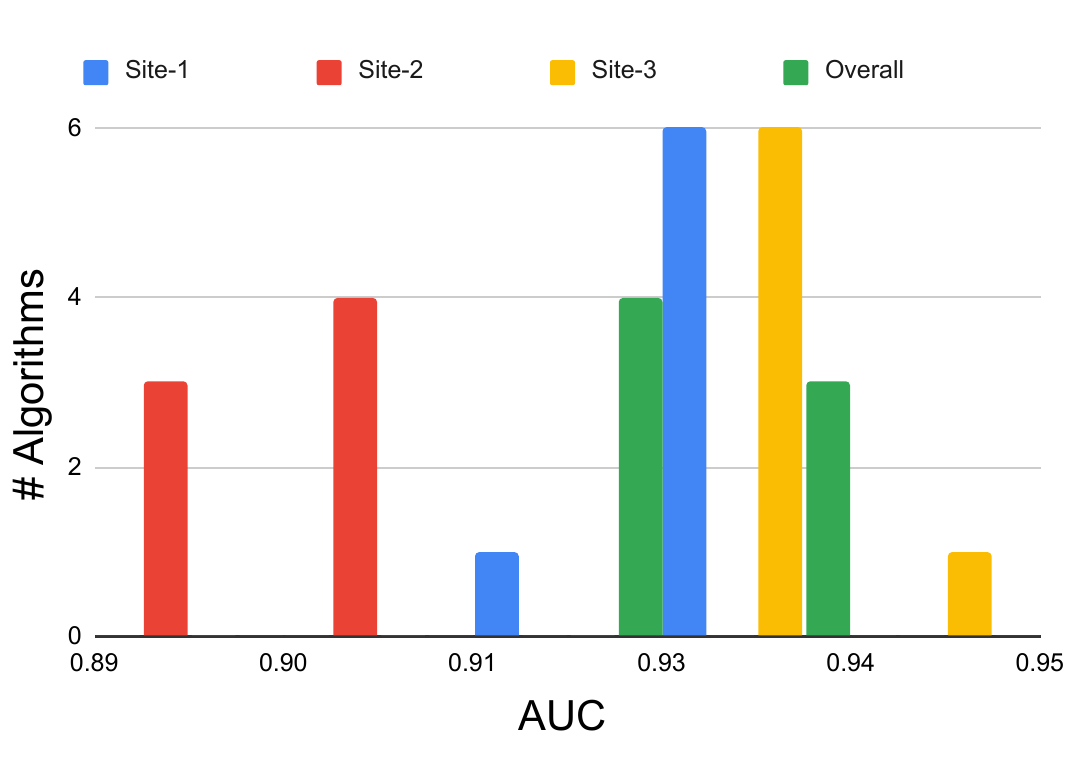}}
    \caption{Site-based metrics in Phase II (Test phase). \label{fig:test_phase_scores}}
\end{figure*}
Figure~\ref{fig:test_phase_scores} shows the average distribution of evaluation metrics across the different sites. \replaced{Specifically, the y axis shows the number of final algorithms to fall into a value range for linear kappa, quadratic kappa, and AUC metrics, shown on the x axis of Figure~\ref{fig:test_phase_scores} a-c respectfully. From the left-most position of the red bars on all plots, one}{One} can observe that the performance of the algorithms is markedly lower on \replaced{site 2}{Site-2}, which included the least training data (as shown in Table~\ref{tab:nr_images}).

We used the statistical ranking toolkit proposed by \citet{Wiesenfarth2021-bj} to analyze the stability of the ranking results. For this evaluation, the algorithms were randomly numbered from \#1 to \#7 and evaluated using the per-image metrics. \deleted{Figure 5 shows bar plots of the per-image distance metrics for each algorithm.}

\begin{figure*}[htbp]
    \centering
    \subfloat[Overall Linear Distance]{\includegraphics[width=0.95\columnwidth]{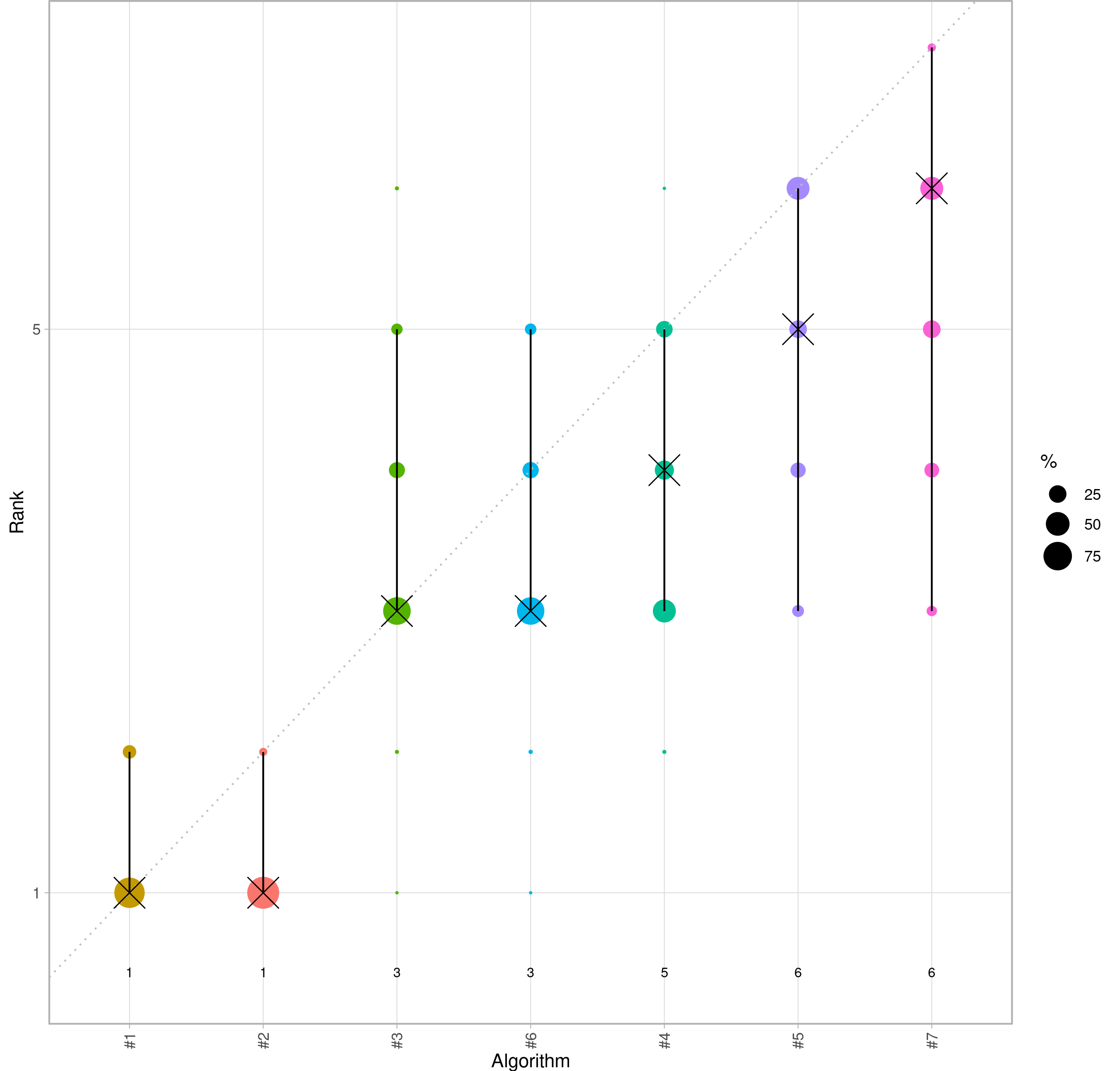}}
    \subfloat[Overall Quadratic Distance]{\includegraphics[width=0.95\columnwidth]{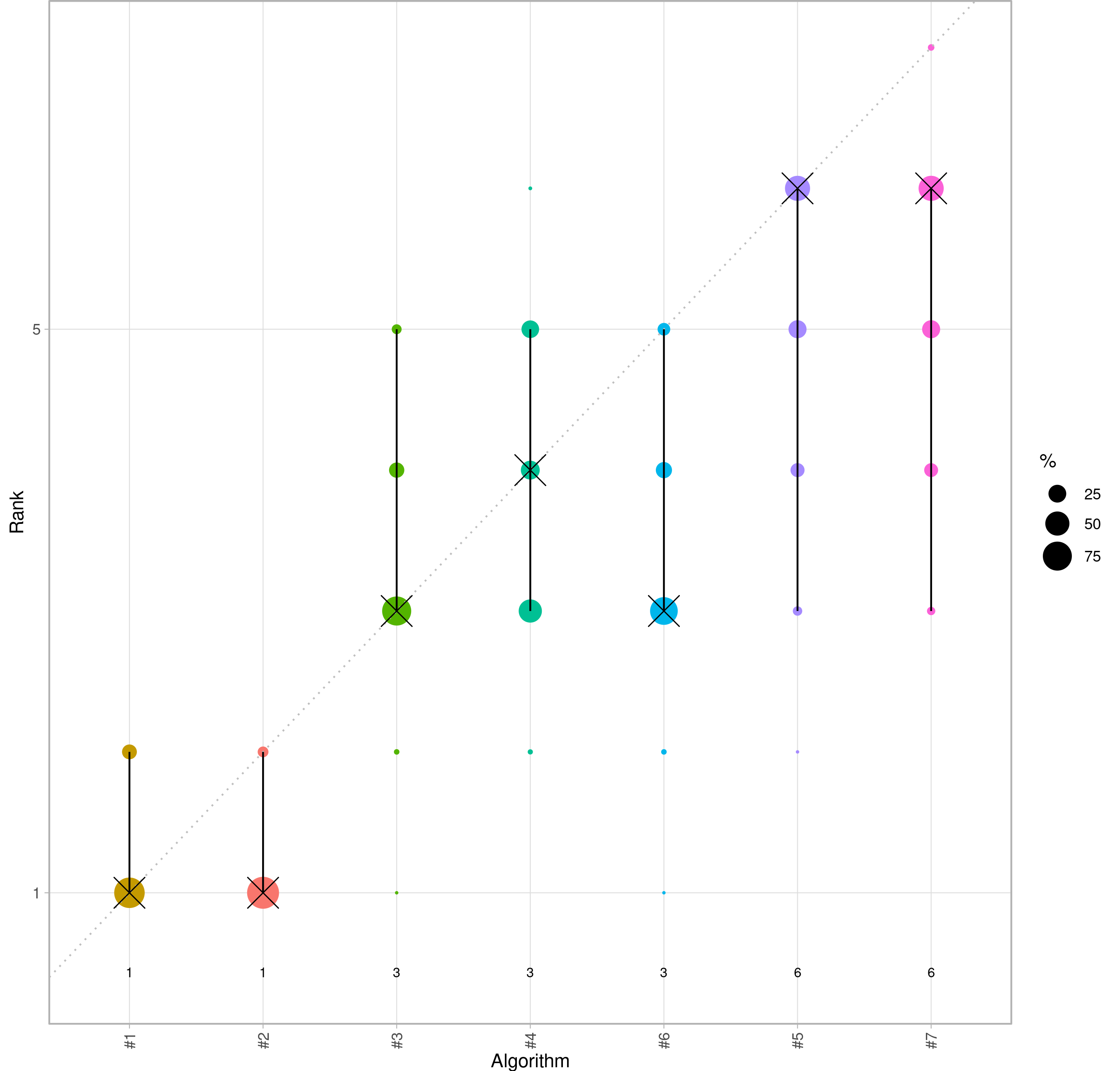}}  
    \caption{Ranking stability using per-image distance metrics in Phase II (Test phase). \label{fig:ranking_stability}}
\end{figure*}
In Fig.~\ref{fig:ranking_stability}, we investigated the ranking stability of the algorithms using the \deleted{per-site and }overall per-image metrics \added{(linear and quadratic)}. The size of each marker at a position represents how often a specific algorithm achieved a certain rank out of 1000 trials using bootstrapping. The black cross shows the median rank for each algorithm, and the black lines show the 95\% range of values from all trials.

\begin{figure*}[htbp]
    \centering
    \subfloat[Overall Linear Distance]{\includegraphics[width=0.75\columnwidth]{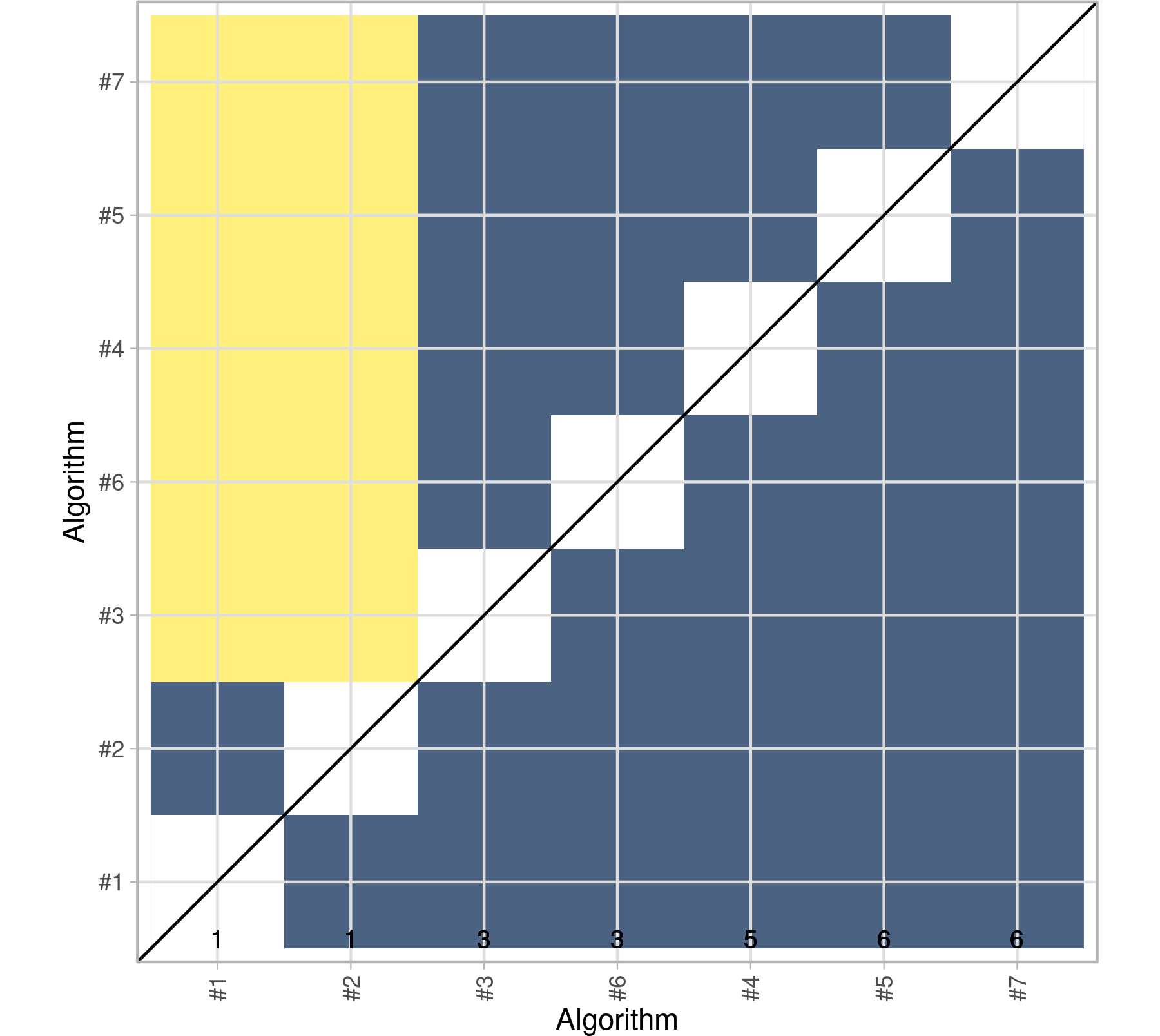}}
    \subfloat[Overall Quadratic Distance]{\includegraphics[width=0.75\columnwidth]{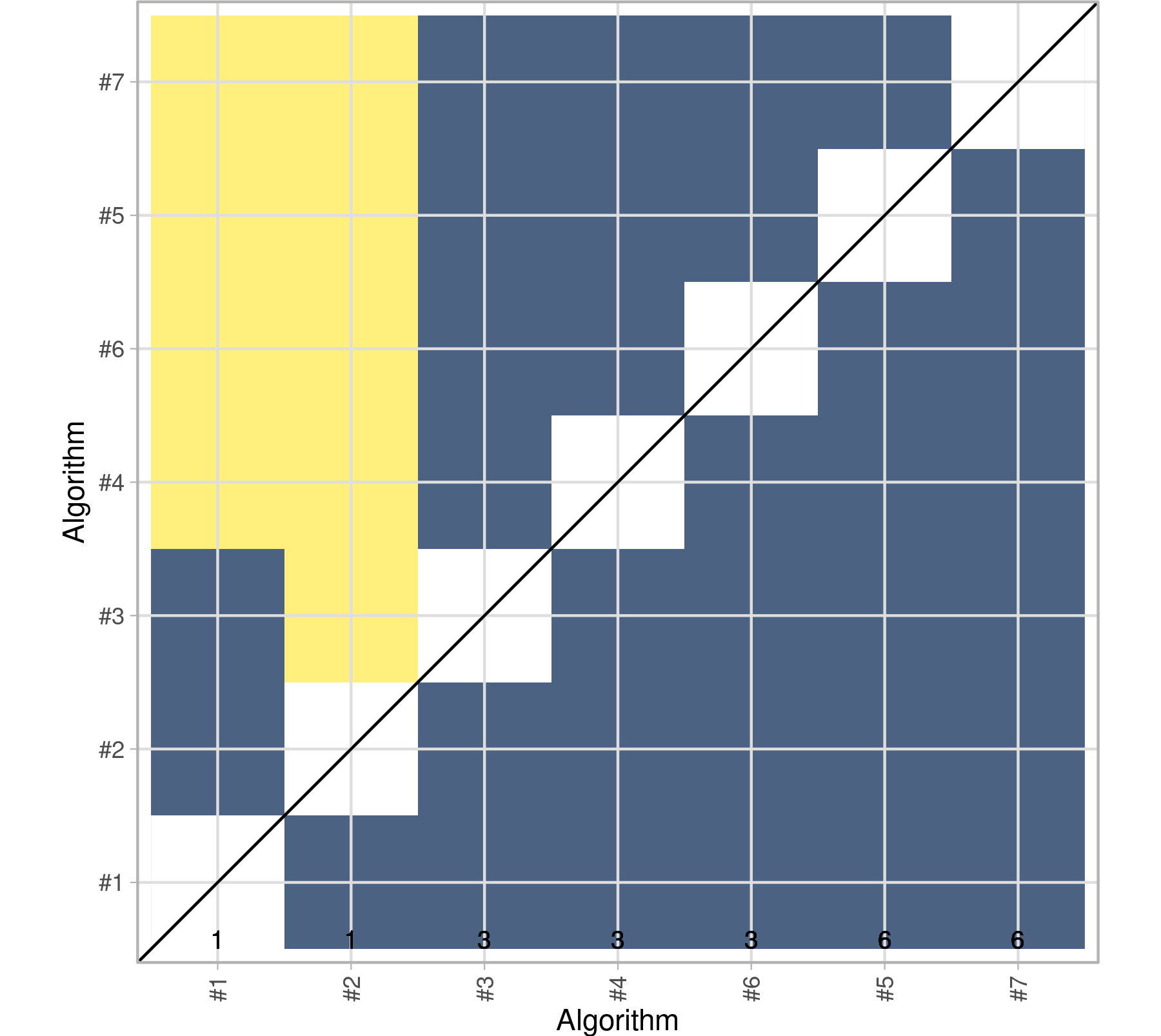}}  
    \caption{Significance maps for overall per-image distance metrics in Phase II (Test phase). \label{fig:significance_maps}}
\end{figure*}

Finally, we compare the statistical significance with which algorithms outperformed other algorithms in different per-image metric tasks. Figure~\ref{fig:significance_maps} shows the significance maps on the overall linear and quadratic per-image metric tasks. Yellow shading indicates the algorithm on the x-axis had significantly better performance than the one on the y-axis. Blue shading indicates no significant difference between the two algorithms.

\replaced{Additional insight into the assessment data utilized in the ranking is given in~\ref{sec:ranking_analysis}.}{To summarize the statistical ranking of the finalist algorithms, we show their ranking stability over all per-image metric tasks in Fig.~\ref{fig:overall_ranking_stability}. Here, the plots show results from multiple tasks separated by algorithm, providing more insight into the assessment data. This can help to better understand the characteristics of each task and the level of uncertainty in ranking the algorithms for each task.}

\subsection{External evaluation}
\begin{table*}[htbp]
    \caption{Comparison of final challenge test ranking and external validation ranking.  \label{tab:leaderboards}}
    \centering\footnotesize
    \subfloat[Final Test Phase Ranking (Phase II)]{
    \begin{tabular}{p{1.5em}p{1.5em}p{3em}p{3em}p{3em}p{3em}p{3em}p{3em}p{3em}p{3em}p{3em}p{3em}p{3em}p{3em}p{3em}}
    \hline
        \textbf{Rank} & \textbf{Algo.} & \textbf{Avg. Rank} & \textbf{Lin. Kappa Site-1} & \textbf{Quad. Kappa Site-1} & \textbf{AUC Site-1} & \textbf{Lin. Kappa Site-2} & \textbf{Quad. Kappa Site-2} & \textbf{AUC Site-2} & \textbf{Lin. Kappa Site-3} & \textbf{Quad. Kappa Site-3} & \textbf{AUC Site-3} & \textbf{Lin. Kappa all} & \textbf{Quad. Kappa all} & \textbf{AUC all} \\ \hline
        \textbf{1st} & \#1 & 1.45 & 0.667 & 0.767 & 0.934 & 0.574 & 0.678 & 0.908 & 0.665 & 0.761 & 0.936 & 0.653 & 0.752 & 0.930 \\ 
        \textbf{2nd} & \#2 & 1.75 & 0.658 & 0.749 & 0.933 & 0.498 & 0.605 & 0.911 & 0.671 & 0.764 & 0.936 & 0.643 & 0.739 & 0.927 \\ 
        \textbf{3rd} & \#3 & 3.15 & 0.605 & 0.711 & 0.929 & 0.493 & 0.607 & 0.905 & 0.671 & 0.766 & 0.939 & 0.626 & 0.729 & 0.928 \\ 
        \textbf{4th} & \#4 & 3.80 & 0.621 & 0.721 & 0.931 & 0.476 & 0.589 & 0.908 & 0.655 & 0.757 & 0.935 & 0.620 & 0.725 & 0.925 \\ 
        \textbf{5th} & \#6 & 4.80 & 0.614 & 0.716 & 0.930 & 0.419 & 0.535 & 0.897 & 0.647 & 0.738 & 0.935 & 0.604 & 0.704 & 0.925 \\ 
        \textbf{6th} & \#5 & 4.85 & 0.627 & 0.728 & 0.924 & 0.421 & 0.533 & 0.894 & 0.646 & 0.747 & 0.938 & 0.608 & 0.713 & 0.922 \\ 
        \textbf{7th} & \#7 & 5.65 & 0.623 & 0.722 & 0.928 & 0.420 & 0.532 & 0.901 & 0.639 & 0.743 & 0.934 & 0.603 & 0.710 & 0.924 \\ \hline
    \end{tabular}
    }\\
    \subfloat[External Validation Ranking]{
    \begin{tabular}{p{1.5em}p{1.5em}p{3em}p{3em}p{3em}p{3em}}
    \toprule
        \textbf{Rank} & \textbf{Algo.} & \textbf{Avg. Rank} & \textbf{Lin. Kappa} & \textbf{Quad. Kappa} & \textbf{AUC} \\ \hline
        \textbf{1st} & \#1 & 1.8 & 0.413 & 0.555 & 0.917 \\ 
        \textbf{2nd} & \#2 & 2.0 & 0.357 & 0.423 & 0.860 \\ 
        \textbf{3rd} & \#3 & 3.0 & 0.350 & 0.427 & 0.851 \\ 
        \textbf{4th} & \#4 & 4.0 & 0.243 & 0.354 & 0.880 \\ 
        \textbf{5th} & \#7 & 4.2 & 0.218 & 0.273 & 0.855 \\ 
        \textbf{6th} & \#6 & 5.4 & 0.172 & 0.253 & 0.853 \\ 
        \textbf{7th} & \#5 & 6.2 & 0.116 & 0.161 & 0.779 \\ 
        \hline
        \textbf{Central} &  & N/A & 0.468 & 0.600 & 0.924 \\
    \bottomrule
    \end{tabular}
    }
\end{table*}

After the challenge concluded, we evaluated the finalist algorithms on an external testing dataset that was not used during the training and test phases of the challenge. Here, we compare the consistency of the algorithm ranking on this external dataset with the Phase II challenge ranking. Table~\ref{tab:leaderboards} shows both the final ranking of the challenge, and the ranking for the same algorithms on the external validation set. While one can observe a noticeable drop in the Kappa and AUC metrics for the external validation, the ranking is relatively consistent with the challenge test ranking. The central DMIST model was also evaluated against the same external testing dataset and results are shown as the last row in Table~\ref{tab:leaderboards}.

\subsection{Demographic Analysis}
\begin{figure*}[htbp]
    \centering
    \includegraphics[width=0.75\textwidth]{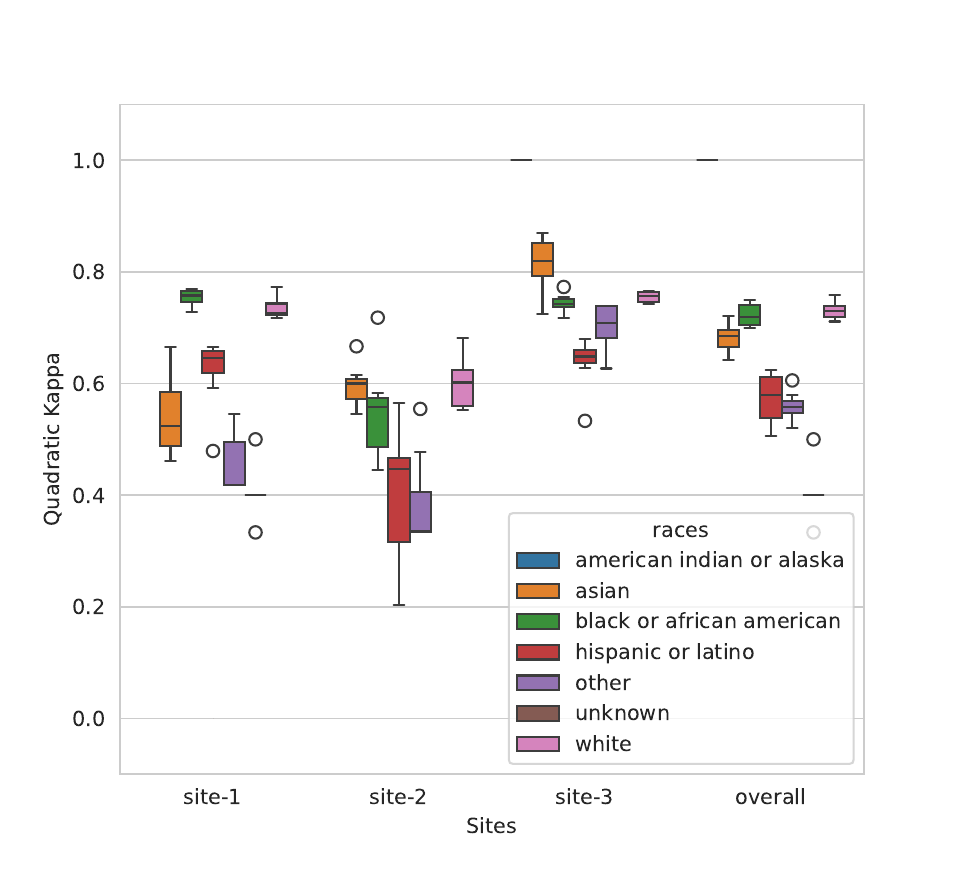}
    \caption{Boxplots showing the quadratic Kappa performance distributions of all finalist algorithms on the Phase II test set. \label{fig:score_race}}
\end{figure*}
\begin{figure}[htbp]
    \centering
    \includegraphics[width=0.95\columnwidth]{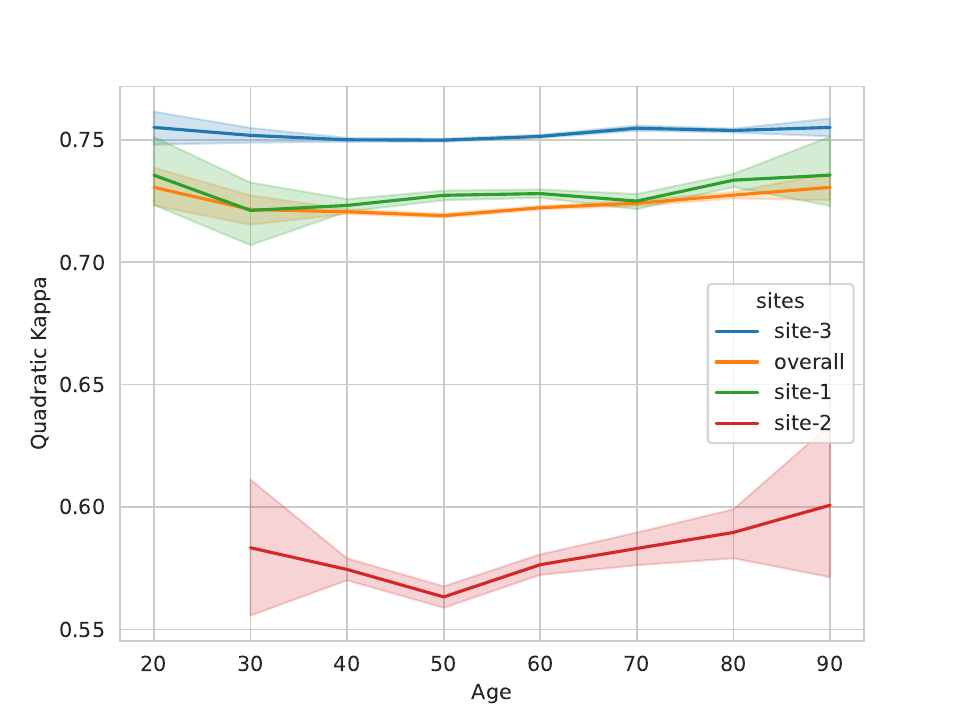}
    \caption{Average quadratic kappa performance of the finalists on the Phase II test set. The shaded area shows the 95\% confidence intervals. \label{fig:score_age}}
\end{figure}

The box plots in Fig.~\ref{fig:score_race} show the distribution of the finalists algorithms across race. Finally, we compared the average performance of the different finalist algorithms across the age of the patients (rounded to the closest 10) in Fig.~\ref{fig:score_age}.

\section{Discussion}
\label{sec:discussion}
The breast density FL challenge attracted world-wide participation and yielded many insights into FL methodology. It also represents a framework for hosting future AI challenges in which data is not distributed to participants and remains behind firewalls. Though conducting the challenge in this way likely required more hands-on debugging from challenge organizers, it made possible the utilization of a rich clinical trial dataset which would ordinarily require explicit IRB approval. 

All final FL algorithms showed worse performance on site 2, the site with the smallest dataset, than \added{the} other \replaced{two}{2} sites. This indicates the global models had better learned features from the more prevalent data types. Interestingly, the use of state-of-the-art FL algorithms such as FedProx~\citep{Li2018-xf} and SCAFFOLD~\citep{Karimireddy2019-bi} did not guarantee higher rank on image-level metrics from site 2. \added{Still, }FedProx was used in the rank 1 algorithm which consistently ranked 1st on site 2 image-level metrics. \added{This can likely be attributed to the reduced model drift when using FedProx as the algorithm introduces an additional penalty term to avoid the models moving to far away from the current global model during local training~\citep{Li2018-xf}}. However,  during Phase I, the rank 5 team found that using FedProx in their implementation resulted in lower overall performance than FedAvg and opted to use FedAvg in Phase II\added{. This might have been caused by a sub-optimal hyperparameter choice}. The rank 2 and 3 algorithms had similar performance on site 2 data, but the rank 2 algorithm used FedAvg while the rank 3 algorithm used SCAFFOLD. \added{SCAFFOLD adds control terms to the local model updates to correct for client drift~\citep{Karimireddy2019-bi}. Given these findings, in this challenge using a real-world data distribution, the aggregation method perhaps proved less relevant to the overall algorithm performances than other methodological choices. We suspect that pre-training and model ensembling played a bigger role in improving the overall robustness of the algorithms.}

While most of the finalist algorithms began FL runs with pretrained model weights, the rank 7 algorithm reported using randomly initialized weights. Using pretrained weights in FL has been shown to enhance performance of global models~\citep{Nguyen2022-kk}. Of the finalist algorithms which used pretrained weights, ranks 2,3,5 and 6 used general purpose pretrained weights. The rank 1 and 4 algorithms used task specific external data for model pretraining. Task specific pre-training may have contributed to the overall performance of these algorithms, but alone did not guarantee a top rank.

The rank 1 and rank 2 algorithms showed similar performance across all metrics, though their approaches were very different. While both algorithms maintained local classifiers at each site, the rank 1 algorithm relied exclusively on the local classifiers for making predictions while the rank 2 algorithm used the local classifiers to constrain updates to a global classifier. Thus, the rank 1 submission did not generate a single complete global model. For validation against the external dataset for that algorithm, the three local site model\replaced{s were treated as an ensemble and the logits}{ outputs} were averaged to obtain a single set of \replaced{probabilities}{metrics}. \added{From there, metrics were computed as per all other submissions.} Though it is well known that local fine tuning often improves AI performance, the lack of a global classifier could pose logistical challenges for AI \added{model} distribution. However, in the case of the rank 1 algorithm, the external validation performance was still highest relative to the other challenge algorithms.

\added{As all but one finalist algorithm produced a single global model, we considered whether ensembling models could be considered unfair. However, the challenge rules did not specifically disallow ensembles of models. Additionally, the stated purpose of this challenge was to allow participants to explore all aspects of FL methodology without constraints, the goal being to find the approach that works best for this FL task. Hence, it was determined that there should be no restrictions against model ensembles.}

Our evaluation of the finalist algorithms on an external testing dataset confirmed a consistent ranking of the top algorithms. This consistency confirms the winning algorithm also generalizes best to this external test set, albeit with a performance drop. Indeed, the ranking order for the first four algorithms stays the same. Only the ranking in 5th to 7th rank changed on this external dataset, further validating the robustness of our ranking results for the challenge. Though the central DMIST model achieved the highest overall performance on the external testing dataset, the central metrics are similar to the rank 1 FL algorithm. 

\added{Aside from capping runtime at 8 hours, computational efficiency was not tracked in this challenge. Unlike some challenges where computational resources are restricted, the participants had full use of all virtual machine resources. That said, in their final algorithm summaries, some participants mentioned encountering issues related to exceeding the allotted 8 hour runtime and lack of sufficient messaging bandwidth between clients and server. During the challenge, challenge organizers actively reported details around each submission failure in order to enable participants to alter their submissions to achieve success. Anecdotally, participants coped with runtime issues by choosing smaller model architectures, reducing the number of training rounds or reducing the number of local epochs between training rounds. The communication bandwidth was imposed at 2 GB by gRPC, the remote procedure call framework used by NVIDIA FLARE. Participants worked around these issues by choosing smaller model architectures. The rank 3 algorithm also used Top-k sparsification to decrease the client-to-server message size}\footnote{\added{Note, the latest 2.4.0 version of NVIDIA FLARE now has a streaming layer that allows message sizes exceeding 2 GB}}\added{.}

The demographic analysis showed that while most algorithms performed relatively comparably across the different race populations and age groups, some markedly lower performances in the latino category can be observed, which is likely attributed to the relatively smaller training data for this race group. We again see a performance drop for most race classes on site 2 compared to sites 1 and 3. The race classes with largest representation (White, Black or African American, and Hispanic or Latino) have stable performance at sites 1 and 3. Some minority race groups like ``american indian or alaska'' have too few samples to draw conclusive results, but we included them for completeness. These findings point to further possibilities for research into fairness and unbiased AI, which could help address these issues. Personalization for specific ethnicities might be another opportunity to address these issues~\citep{Zhong2022-fk}.

\section{Conclusion}
In conclusion, the breast density FL challenge provided insights into FL methodology and a framework for hosting future distributed AI challenges without access to the data for participants, allowing for a fair comparison between the algorithms. Importantly, we have demonstrated how FL algorithm development can be completely decoupled from access to patient data itself, an important proof-of-concept for medical AI that extends beyond competitions. Although the FL algorithms generalized well on the challenge testing set, the drops in performance on external data and minority groups further highlight the need for research into fairness and unbiased and personalized AI models. We hope that the challenge's success encourages hosting future AI challenges for fair evaluation of state-of-the-art FL methods.

\section*{Acknowledgments}
The environment for this challenge was supported by The Microsoft AI for Health grant. 
M.E. K.M. are partially funded by the EACEA Erasmus Mundus grant, R.M. acknowledges the funding from the Spanish Science and Education Ministry, grant no. PID2021-123390OB-C21.
R.Z. is partially supported by Wu Wen Jun Honorary Doctoral Scholarship, AI Institute, Shanghai Jiao Tong University.

Funding for challenge support, via a subcontract, was provided by Leidos Biomedical Research, Inc which provides operational and technical support to the Frederick National Laboratory for Cancer Research, a Federally Funded Research and Development Center (FFRDC). Operational and technical support involves the management and execution of projects sponsored by the National Cancer Institute (NCI) Center for Biomedical Informatics and Information Technology (CBIIT). The Medical Imaging Challenge Infrastructure (MedICI) Platform was developed to support imaging, digital pathology, and genomics challenges. This system was built on a framework of open-source projects and was used in the breast density classification challenge.

\bibliographystyle{model2-names.bst}\biboptions{authoryear}
\bibliography{refs}

\clearpage
\newpage
\onecolumn
\appendix
\section{\deleted{Supplementary Material: }Descriptions of Finalist Algorithms}
\label{sec:abstracts}
\begin{center}

\begin{minipage}{0.85\textwidth}
\begin{center}
\textbf{\underline{1$^{st}$ Rank} (Algo. \#1)\\
Federated Breast Density Estimation with Deep Ordinal Regression}\\
Marawan Elbatel$^\ast$, Kaouther Mouheb$^\ast$, Robert Marti\\
\textit{Computer Vision and Robotics Institute, University of Girona.}\\
$^\ast$Equal contribution
\end{center}
Inspired by the fact that local learning matters in federated learning (FL)~\citep{Chen2021-js,Mendieta2022-tk}, we focus on local client training in our submission. First, we pre-process the non-IID input mammograms across different sites using histogram equalization following a previous observation~\citep{Roth2020-vb}. Although this equalization can result in a loss of intensity information, such harmonization ensured consistent learning among different sites in FL. As the relative ordering between breast density classes is significant, we propose to leverage such information and formulate the problem as an ordinal classification task. Thus, we add a rank-consistent layer~\citep{cao2020rank} on top of the feature encoder, ResNet50, to perform the deep ordinal classification. For faster convergence in the FL setting, we initialize the local clients’ models with weights extracted from an upstream object detection task on the DBTeX challenge~\citep{Konz2023-so}. We utilize our local training procedure in the NVflare framework with FedAvg~\citep{Konz2023-so,McMahan2017-tc}, FedProx~\citep{Mendieta2022-tk,Li2018-xf}, FedAlign\cite{Mendieta2022-tk}, and MOON~\citep{Li2021-pw}. Experimental results showed that FedProx yielded the highest performance over the validation set. Finally, we train all clients with FedProx with one local epoch for 100 communication rounds.
\end{minipage}

\begin{minipage}{0.85\textwidth}
\begin{center}
\textbf{\underline{2$^{nd}$ Rank} (Algo. \#2)\\
A Model Mixture Framework for Personalized Federated Learning in Breast Density Estimation}\\
Ruipeng Zhang$^{1,2}$, Yao Zhang$^{2}$, Yanfeng Wang$^{1,2}$\\
\textit{$^{1}$Cooperative Medianet Innovation Center, Shanghai Jiao Tong University\\
$^{2}$Shanghai AI Laboratory}
\end{center}
The Breast Density FL challenge aims at developing generalizable models from distributed datasets for breast density estimation. While federated learning methods, e.g., FedAvg, enable learning from multi-site datasets without data sharing, the heterogenous data still remains a great challenge to current methods. The data heterogeneity lies in two aspects: the variance of data distribution and the imbalance of classes among multi-site datasets. Therefore, we are motivated to avoid the global model being biased by the dominant site while keeping the local models powerful for their corresponding sites.

To this end, we propose a personalized federated learning framework based on a model mixture. In our framework, each site has its own local model that consists of a shared feature extractor and two classifiers: one local and one global. Both local and global classifiers learn from the local dataset. In contrast, only the parameters of the global classifier and those of the feature extractor are aggregated and updated to form the global model during the federated learning process. For balanced training, we apply a cross-entropy loss to both the sum of two classifiers’ predictions and another solely for only the global classifier.

Specifically, we employ a ResNet50 pre-trained on ImageNet as our choice of feature extractor. The framework is trained for 80 epochs, and the parameter aggregation is performed every 2 epochs. In the evaluation phase, our method achieves generalized results on the less represented sites (sites 1 and 2) and also keeps a comparable performance on the most represented site (site 3).
\end{minipage}

\begin{minipage}{0.85\textwidth}
\begin{center}
\textbf{\underline{3$^{rd}$ Rank} (Algo. \#3)\\
Simple yet Effective Federated Learning Workflow for Non-iid Breast Density Classification with CommunicationLimitation}\\
Yaojun Hu$^{1}$, Haochao Ying$^{1,2}$, Yuyang Xu$^{1,3}$\\
\textit{$^{1}$Real Doctor AI Research Centre, Zhejiang University, China\\
$^{2}$School of Public Health, Zhejiang University, China\\
$^{3}$College of Computer Science and Technology, Zhejiang University, China}
\end{center}
Accurately estimating breast density is of significant value in early detection of masked breast cancer. However, in this competition, we observe two important issues. First, the multi-institutional competition dataset is not independently and identically distributed. Second, Competition training and testing environments have communication limitations. To this end, we employ federated averaging workflow, exactly SCAFFOLD algorithm, for data heterogeneity issue, along with Top-k gradient sparse for communication cost. Basically, DenseNet-121 model is utilized as the base model. Finally, we rank third with the average performance of 0.626 in linear kappa, 0.729 in quad kappa, and 0.928 in AUC at three testing institutional data, respectively.
\end{minipage}

\begin{minipage}{0.85\textwidth}
\begin{center}
\textbf{\underline{4$^{th}$ Rank} (Algo. \#4)\\
Breast Density Classification Using Federate Learning Using Pretrained Breast Density Classifiers}\\
Conrad Testagrose$^{1}$, Mutlu Demirer$^{2}$, Vikash Gupta$^{2}$, Xudong Liu$^{1}$, Barbaros S. Erdal$^{2}$, Richard D. White$^{2}$\\
\textit{$^{1}$University of North Florida College of Computing, Jacksonville, United States\\
$^{2}$Mayo Clinic Florida Radiology - CAII, Jacksonville, United States}
\end{center}
We used transfer learning for performing this classification challenge. Inception-V3~\citep{Szegedy2016-mq} is used as the base network with weights from ImageNet~\citep{Alex_Krizhevsky_Ilya_Sutskever_Geoffrey_E_Hinton2017-cx} as initialization. The images were resized to a resolution of 299 x 299 and the single grayscale channel was converted to a three-channel image (similar to color images in the ImageNet~\citep{Alex_Krizhevsky_Ilya_Sutskever_Geoffrey_E_Hinton2017-cx} dataset). This model was further fine-tuned using data from the Mayo Clinic Health system. The Mayo Clinic data used was comprised of 118,459 total mammogram images. Of these images, 55826 were C-View images (Hologic, Marlborough, MA, https://www.hologic.com/), 49040 were Full Field Digital Mammography images, and 13,593 were Tomosynthesis Projection images. This model was then further fine-tuned using the training data provided for the challenge. Data preprocessing techniques similar to that used on the Mayo Clinic data were used on the challenge training data. Data augmentation techniques such as image rotation, flipping, and normalization were performed on the images in this preprocessing step. The submitted model was trained using a learning rate of 0.001 with a batch size 16 and the Adam Optimizer~\citep{Kingma2014-fp}. The training pipeline tools were built using PyTorch~\citep{Paszke2019-vz}, NVFlare~\citep{Roth2022-ru}, and the MONAI~\citep{Cardoso2022-uy} machine learning toolkit. 
\end{minipage}

\begin{minipage}{0.85\textwidth}
\begin{center}
\textbf{\underline{5$^{th}$ Rank} (Algo. \#6)\\
FedAvg-ed ResNet18 for Breast Density Classification}\\
Ünal Akünal$^\ast$, Markus Bujotzek$^\ast$, Klaus H. Maier-Hein\\
\textit{Division of Medical image Computing, German Cancer Research Center, Heidelberg, Germany}\\
$^\ast$Equal contribution
\end{center}
We recognized the similarity between the conducted Federated Learning challenge for breast density classification and the work by \citet{Roth2020-vb}. Inspired by it, we started the federated training with a FedAvg aggregation strategy for 150 federated communication rounds while performing 1 local epoch per round. As a model, DenseNet121 was trained via a CE-loss and Adam optimizer (lr=1e-4) with a batch size of 64. This SOTA setup failed for the given challenge due to the boundary condition that a training should not last longer than 8 hours. Moreover, we found in further experiments that models larger than ResNet50 couldn’t be handled by the provided training infrastructure due to the computational constraints on message size. To reduce both training duration and model size, we selected ResNet18 as a model trained with a batch size of 64 via CE-loss and SGD optimizer (lr=2e-3, momentum=0.9) to achieve faster convergence in the limited training time. We stuck to FedAvg as an aggregation strategy, but reduced the number of federated communication rounds to 100 while still performing 1 local epoch per communication round. With this setup we also achieved the best training results measured by the applied metrics and therefore submitted this setup for testing. We finalized our investigations by modifying FedAvg to FedProx method to tackle a potentially present domain shift between the dataset splits. However, since our FedProx setup performed worse than the previously described setup, we concluded that the present domain shift between the data splits is not as heterogenous in the sense that the proximal term in FedProx does not contribute to but hinders the performance of the global model compared to the previous FedAvg experiments.
\end{minipage}

\begin{minipage}{0.85\textwidth}
\begin{center}
\textbf{\underline{6$^{th}$ Rank} (Algo. \#5)\\
Untitled submission}\\
Ghanem Bahrini\\
\textit{Université de Carthage, Tunis, Tunesia}\\
\end{center}
[Note from challenge organizers] The author did not submit a description of this method. However, we include the result for completeness as it was used in the final ranking of the challenge. The method is based on the baseline implementation provided by the challenge organizers with only a slight modification, i.e., using ResNet50 instead of the baseline ResNet18.
\end{minipage}

\begin{minipage}{0.85\textwidth}
\begin{center}
\textbf{\underline{7$^{th}$ Rank} (Algo. \#7)\\
Fine-tuned FedAvg for Federated Breast Density Estimation}\\
Yi Qin, Xiaomeng Li\\
\textit{Electronic and Computer Engineering, Hong Kong University of Science and Technology}
\end{center}
Accurate breast density estimation is essential in early breast cancer diagnosis. Federated learning provides privacy-preserving solutions for precisely estimating the density from multiple isolated data sources. In this challenge, we finetuned the FedAvg algorithm to provide a solution for the federated breast density estimation task. We used the local data and their partition from the organizer as our client's input. First, each client trained their local ResNet-56 model to minimize the cross entropy loss for five epochs. Then, clients uploaded the parameters of their local encoder to the server. After this, the server aggregated a new global model by weighted averaging the received clients' model weight. The weight is determined by the proportion of the local client's data size to the total data amount. Finally, the server updated the local models by distributing the global model to local clients. We deployed a grid search to derive the optimal hyperparameters. The submitted results achieved AUC all 0.899, Lin. Kappa all 0.59, and Quad. Kappa all 0.698.
\end{minipage}

\end{center}

\section{\added{Additional Ranking Analysis}}
\label{sec:ranking_analysis}

\begin{figure*}[htbp]
    \centering
    \includegraphics[width=0.95\textwidth]{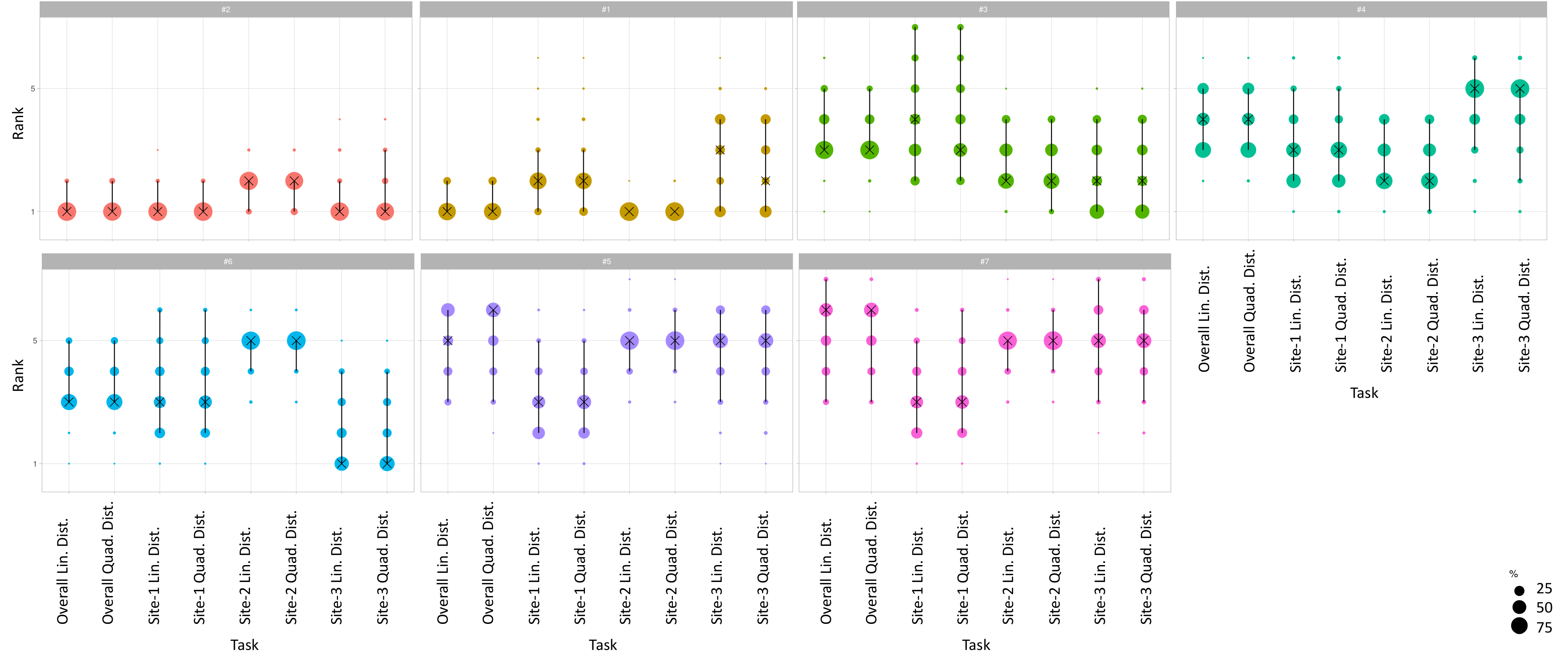}
    \caption{Overall Ranking stability across different metrics and challenge tasks. Figures are ordered from first to last rank, going from left to right, top to bottom. \label{fig:overall_ranking_stability}}
\end{figure*}

\added{To summarize the statistical ranking of the finalist algorithms, we show their ranking stability over all per-image metric tasks in Fig.~\ref{fig:overall_ranking_stability}. Here, the plots show results from multiple tasks separated by algorithm, providing more insight into the assessment data. This can help to better understand the characteristics of each task and the level of uncertainty in ranking the algorithms for each task.}

\end{document}